\documentclass[twocolumn,showpacs,preprintnumbers,amsmath,amssymb]{revtex4}

\usepackage{graphicx}
\usepackage{dcolumn}
\usepackage{bm}
\begin{document}


\title{Temperature Dependence of the Band Gap of Semiconducting Carbon Nanotubes}

\author{Rodrigo B. Capaz$^{1,2,3}$, Catalin D. Spataru$^{2,3}$, Paul Tangney$^{2,3}$, 
Marvin L. Cohen$^{2,3}$, and Steven G. Louie$^{2,3}$}

\affiliation{$^1$ Instituto de F\'\i sica, Universidade Federal do Rio de Janeiro, Caixa Postal 68528, Rio de Janeiro, RJ 21941-972, Brazil \\
$^2$ Department of Physics, University of California at Berkeley, Berkeley, CA 94720 \\
$^3$ Materials Science Division, Lawrence Berkeley National Laboratory, Berkeley, CA 94720
}

\date{\today}

\begin{abstract}
The temperature dependence of the band gap of semiconducting single-wall carbon nanotubes 
(SWNTs) is calculated by direct evaluation of electron-phonon couplings within a ``frozen-phonon''
scheme. An interesting diameter and chirality dependence of $E_g(T)$ is obtained, including non-monotonic
behavior for certain tubes and distinct ``family'' behavior. These results
are traced to a strong and complex coupling between band-edge states and the lowest-energy
optical phonon modes in SWNTs. The $E_g(T)$ curves are modeled by an analytic function 
with diameter and chirality dependent parameters; these provide a valuable guide for
systematic estimates of $E_g(T)$ for any given SWNT. Magnitudes of the temperature shifts
at 300 K are smaller than 12 meV and should not affect $(n,m)$ assignments based on optical 
measurements.
\end{abstract}

\pacs{73.22.-f, 63.22.+m, 71.38.-k}
\maketitle

The temperature dependence of the band gap ($E_g$) is one of the fundamental
signatures of a semiconductor, providing important insights into
the nature and strength of electron-phonon ({\it e-p}) interactions. The first 
measurements of $E_g(T)$ date from the dawn of the semiconductor 
era \cite{becker}. Typically, $E_g(T)$ curves show a monotonic decrease with 
temperature that is non-linear at low $T$ and linear at sufficiently high $T$
\cite{madelung,cardona2}. 

Semiconducting carbon nanotubes are relatively novel semiconductor
materials \cite{iijima}, with a variety of potential applications. 
Despite intensive research since their discovery, it has
only recently become possible to perform measurements of the optical gap 
in individual single-wall carbon nanotubes (SWNTs) 
\cite{jorio,li,oconnell,bachilo,hartschuh,lefebvre1}.
Such measurements, combined with information from vibrational spectroscopy, 
provide a route to $(n,m)$ assignment of SWNTs \cite{jorio,bachilo,hartschuh}. 
Understanding $E_g(T)$ for nanotubes is extremely important in this context, since 
experiments are usually performed at room-temperature and $(n,m)$
assignments are often guided by comparisons between observed optical transition
patterns and the corresponding predictions from calculations at $T = 0$ K. Moreover,
the $E_g(T)$ signature could provide extra information for those assignments.

Although it has been demonstrated that many-body 
quasiparticle and excitonic effects are crucial for the correct description of the
photoexcited states and for a quantitative 
understanding of such optical measurements \cite{spataru}, the single-particle 
gap is still a fundamental quantity because: (i) it is the starting point for more 
elaborate descriptions and (ii) trends in the single-particle gap 
are often preserved by such refinements. Therefore, this work is devoted to
describing the temperature dependence of the single-particle band gap
of semiconducting SWNTs. From the calculated results for 18 different SWNTs, 
a complex dependence of $E_g(T)$ on chirality and diameter is found, with an
unsual {\it non-monotonic} behavior for certain classes of tubes. This
behavior arises from the differences in sign of the {\it e-p} coupling associated 
to low-energy optical phonons. A model relation describing  $E_g(T)$ for 
any given SWNT, as a function of diameter and chirality, is proposed. 

The temperature dependence of $E_g$ at constant pressure can be separated into
harmonic and anharmonic contributions: 
$(\partial E_g/\partial T)_P = (\partial E_g/\partial T)_{har} + (\partial E_g/\partial T)_{anh}$.
The harmonic term arises from the {\it e-p} interaction evaluated at the ground-state 
geometry. The anharmonic 
term is due to thermal expansion. The harmonic term is usually the more difficult one to evaluate and 
it has challenged theorists for many years \cite{fan,antoncik,allen1,allen2,allen3}.
We follow closely the formulation of Allen, Heine and Cardona 
\cite{allen1,allen2,allen3} to calculate $E_g(T)$ for semiconducting SWNTs. 
In the spirit of the adiabatic approximation \cite{allen1}, we write the shift
of an electronic eigenenergy $\Delta E_{n,{\bf k}}$ of band $n$ and wavevector
${\bf k}$ due to static atomic displacements from equilibrium 
as a second-order Taylor expansion: 
\begin{equation}
\label{eq.1}
\Delta E_{n,{\bf k}} = {\bf u} \cdot \nabla E_{n,{\bf k}} + {\frac{1}{2}}{\bf u}\cdot {\sf D} \cdot {\bf u} %
,
\end{equation}
where ${\bf u}$ is a $3N$-coordinate displacement vector and $\sf D$ is the corresponding
($3N \times 3N$) Hessian matrix ($N$ is the number of atoms). As usual, we 
express the displacements as a sum over normal modes \cite{maradudin}:
\begin{equation}
\label{eq.2}
u(\alpha,\kappa) = \sum _j \left({\frac {\hbar}{2M_{\kappa}\omega_j}}\right)^{1/2} \varepsilon_j(\alpha,\kappa) (a{^\dagger}_j + a_j),
\end{equation}
where $u(\alpha,\kappa)$ is the displacement along direction $\alpha$ of atom
$\kappa$ in the unit cell, with mass $M_\kappa$. The electron-phonon couplings will be
directly evaluated using the ``frozen-phonon'' scheme \cite{lam}, therefore the sum 
only runs over zone-center modes $j$, with frequency $\omega_j$. 
In this approach, a sufficiently large supercell (equivalent
to a fine Brillouin Zone sampling) will
be needed to achieve convergence. The $a{^\dagger}_j$ and $a_j$ are creation
and destruction operators for phonons, and $\varepsilon_j(\alpha,\kappa)$ are 
components of the properly normalized polarization vectors.

We now substitute Eq.~(\ref{eq.2}) into Eq.~(\ref{eq.1}) and perform a thermal
average \cite{maradudin}. In the harmonic approximation, the linear term in
${\bf u}$ vanishes and the final result is analogous to Ref.\cite{allen2}'s:
\begin{equation}
\label{eq.3}
\Delta E_{n,{\bf k}} = \sum _j {\frac {\partial E_{n,{\bf k}}}{\partial n_j}}(n_j+{\frac {1}{2}}),
\end{equation}
where $n_j=(e^{\beta \hbar \omega_j}-1)^{-1}$ is the Bose-Einstein occupation
number of phonon mode $j$ and the {\it e-p} coupling coefficient 
$\partial E_{n,{\bf k}}/\partial n_j$ is given by
\begin{equation}
\label{eq.4}
{\frac {\partial E_{n,{\bf k}}}{\partial n_j}}={\frac {1}{2}}{\bf x}_j\cdot {\sf D} \cdot {\bf x}_j,
\end{equation}
where $x_j(\alpha,\kappa)=(\hbar/M_{\kappa}\omega_j)^{1/2} \varepsilon_j(\alpha,\kappa)$ are 
``frozen-phonon'' displacements. From
Eq. (\ref{eq.1}), this is simply the quadratic contribution to 
$\Delta E_{n,{\bf k}}$ when the atoms are displaced along a certain 
frozen-phonon ${\bf x}_j$. So, in practice,
$\partial E_{n,{\bf k}}/\partial n_j$ is calculated by performing 
electronic structure calculations for $\pm {\bf x}_j$ and by averaging
the obtained energy shifts so as to eliminate the linear term.

Structural relaxations and phonon calculations are performed using
the extended Tersoff-Brenner interatomic potential \cite{brenner}. 
Electronic structure 
is calculated using a multiple-neighbor, non-orthogonal tight-binding method 
\cite{hamada,okada}. Calculations are performed for 18 different semiconducting
SWNT's with diameters $d=7.6-13.5$ {\AA } and spanning the entire range of 
chiralities: (6,5), (7,5), (7,6), (8,6), (8,3), (8,4), (9,4), (9,5), (9,1), 
(11,1), (10,2), (12,2), (10,0), (11,0), (13,0), (14,0), (16,0) and (17,0).\cite{fn2} 

\begin{figure}
\includegraphics[width=2.7in]{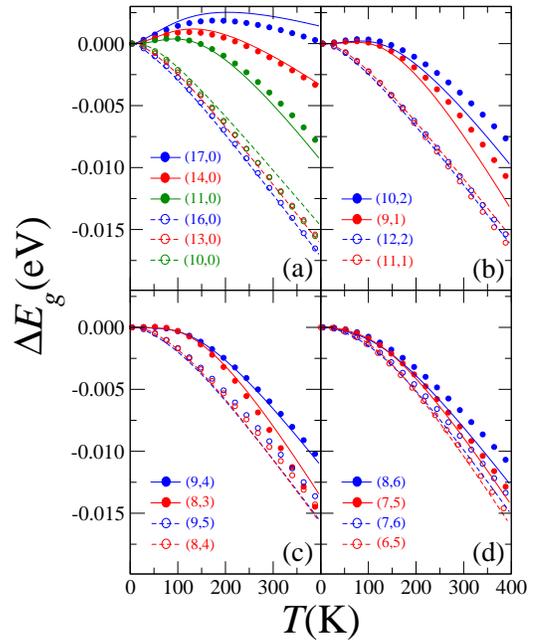}
\caption{Calculated and fitted $\Delta E_g(T)$. Full dots and lines
correspond to $\nu=2$ SWNTs, whereas open dots and dashed lines correspond to $\nu=1$ SWNTs.}
\label{fig.1}
\end{figure}

Fig. \ref{fig.1} shows the calculated (dots) values of 
$\Delta E_g (T)= E_g (T) - E_g (0)$ for all the 18 SWNTs studied. The lines are best fits
to our model relation of Eq. (\ref{eq.6}), to be discussed below. For clarity, results are grouped
into four panels according to similar values of the chiral angle $\theta$. 
The temperature dependence of the band gap is
relatively small compared to bulk semiconductors: 
From the set of calculated SWNTs, the largest value of $E_g(0)-E_g(300K)$ 
is 12 meV for the (16,0) tube, with 
$(dE_g/dT)_{300 K}=-5.2\times 10^{-2}$ meV/K. A strong and 
apparently complicated chirality and diameter dependence emerges: SWNTs
with $\nu=(n-m) \mod 3=2$ (full dots) have, in general, smaller gap shifts 
than $\nu=1$ tubes (open dots). 
Most interestingly, some $\nu=2$ SWNTs with small chiral angles show a 
{\it non-monotonic} gap variation with temperature that is positive for small $T$ 
and negative for larger $T$. Some of these trends, in particular the
overall magnitude of the shifts and their $\nu$-oscillations, are observed in 
recent photoluminescence measurements in suspended SWNTs \cite{lefebvre2}.

It is challenging to explain this interesting  and complex behavior within 
a unified framework. The first step is to analyze the contributions
to $E_g(T)$ from the different phonon modes. This information is 
contained in the {\it e-p} spectral function for the gap, 
$g^2F$, defined as 
\cite{allen2}
\begin{eqnarray}
\label{eq.5}
g^2F(\Omega)=\sum_j{\frac {\partial E_g}{\partial n_j}}\delta(\Omega-\omega_j).  %
\end{eqnarray}
Note that $\partial E_g/\partial n_j$, defined as the difference between coupling coefficients 
for the two band-edge states (from Eq. (\ref{eq.4})), can be positive or negative. 

\begin{figure}
\includegraphics*[width=2.9in]{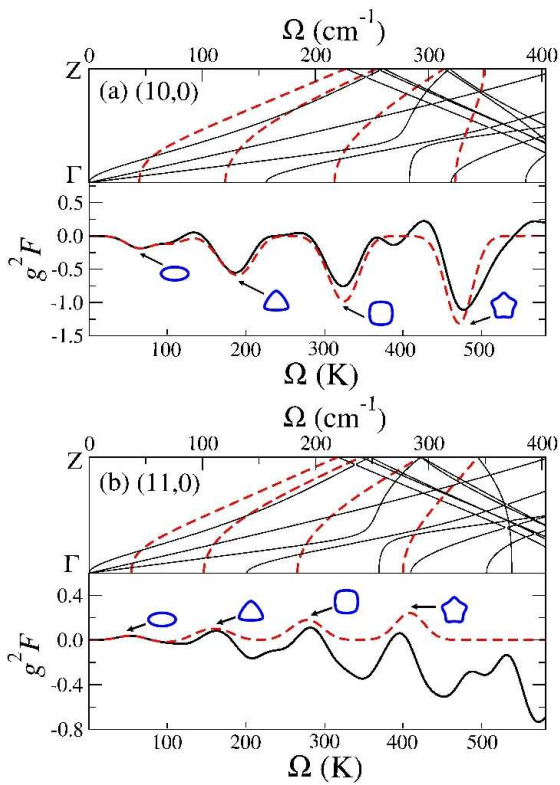}
\caption{Dimensionless $g^2F$ (lower panels) and phonon spectra (upper panels) for 
(a) (10,0) and (b) (11,0) SWNTs. Lower panels: Full black lines are the total $g^2F$ and red 
dashed lines are the contributions from the shape-deformation modes (SDMs) near $\Gamma$. 
The $g^2F$ is obtained from Gaussian broadening ($\sigma = 23$ K) of individual phonon contributions.
Shapes for each SDM are also shown in the figure. Upper panels: Phonon frequencies ($\Omega$) 
measured in both cm$^{-1}$ and K. SDM branches are highlighted in red dashed lines.
}
\label{fig.2}
\end{figure}

In Fig. \ref{fig.2}(a) and (b) (lower panels) we plot $g^2F$ for the (10,0) and (11,0) 
SWNTs, which are prototype examples of $\nu=1$ and $\nu=2$ tubes, respectively. The plots are
restricted to the low-energy phonon branches (shown in the upper panels), the relevant 
ones to describe $E_g(T)$ for $T<400$ K. We notice that $g^2F$ 
(full black line) is highly structured, reflecting the complexity of the phonon 
dispersion of these materials, even at low energies. In particular, we find that the 
low-energy optical ``shape-deformation'' modes (SDMs)
near $\Gamma$ provide the most important contributions to $g^2F$ at low energies,
shown by the red dashed lines. This family of modes is derived, in a zone-folding 
scheme, from the out-of-plane transverse-acoustical (ZA) branch of graphene and 
they deform the circular cross-section of the tubes into a sequence of shapes: 
ellipse (the so-called ``squashing mode''), triangle, square, pentagon, etc. 
For the (10,0) tube, the contribution to $g^2F$ from these modes is strongly 
negative and dominates the full $g^2F$ up to phonon energies equivalent to 500 K.
For the (11,0), this contibution is positive and smaller, but still dominates
$g^2F$ for $T < 200$ K. Beyond that temperature, the negative contributions from 
other modes start to become important. This competition explains the non-monotonic 
behavior and smaller magnitudes of $E_g(T)$ shifts for $\nu=2$ SWNTs. 

Approximate model relations for $E_g(T)$ are common in semiconductor physics 
\cite{passler}. Despite not being exact, they are useful tools 
for a quick assessment of $E_g(T)$ for a given material. 
We propose a model relation for $E_g(T)$ of semiconducting SWNTs in the temperature
range of $T < 400$ K as a two-phonon Vi\~na model \cite{vina}:
\begin{equation}
\label{eq.6}
\Delta E_g(T) = {\frac {\alpha_1 \Theta_1}{e^{\Theta_1/T}-1}} + {\frac {\alpha_2 \Theta_2}{e^{\Theta_2/T}-1}},
\end{equation}
where $\Theta_1$ and $\Theta_2$ ($\Theta_1 < \Theta_2$) are effective temperatures for the 
two ``average phonons'' and $\alpha_j\Theta_j=\partial E_g/\partial n_j$ are their effective 
{\it e-p} coupling coefficients. The resulting fits are shown
in Fig. \ref{fig.1} as full ($\nu=2$) and dashed ($\nu=1$) lines. Note that this model is equivalent to
replacing the highly structured spectral functions shown in Fig. \ref{fig.2} by only two
delta functions for each tube. Therefore, the resulting fits will necessarily be approximate. 
Nonetheless, as we see from Fig.\ref{fig.1}, the largest deviation from numerical and model results is 
only 2 meV in the whole $T<400$ K range. More importantly, as we shall see below, we are 
able to obtain a completely coherent and physically justified dependence of the 
parameters $\alpha_j$ and $\Theta_j$ on the SWNT's diameter and chirality.

We start by considering the parameter $\Theta_1$, the effective frequency for the lowest 
energy phonon modes. These modes dictate the low-temperature ($T\lesssim 100$ K) behavior
of $E_g(T)$. These are the SDMs, derived from the ZA branch of graphene.
In graphene, this branch has a quadratic dispersion at low energies and momenta. Therefore,
within a zone-folding description, we propose that the energies of these modes (and therefore
$\Theta_1$) should scale as the inverse square of the nanotube diameter $d$ (in
dimensionless units, $d=\sqrt{n^2+m^2+nm}$):
\begin{equation}
\label{eq.7}
\Theta_1 = {\frac {A}{d^2}}.
\end{equation}
Therefore, $\Theta_1$ depends only on diameter and not on chirality.

The remaining parameters ($\alpha_1$, $\Theta_2$ and $\alpha_2$) depend on chirality. 
The task of finding this dependence is enormously facilitated by the following $ansatz$: 
Dependence on chirality can be expressed as polynomial expansions $f(\xi)$ of a 
``chirality variable'' $\xi = (-1)^\nu \cos(3\theta)$. This $ansatz$ is reminiscent of 
trigonal warping effects in graphite \cite{saito,reich} and
is justified {\it a posteriori} by our numerical results. We consider up to
2nd-order terms:
\begin{equation}
\label{eq.9}
f_\eta(\xi) = \gamma_1^\eta\xi + \gamma_2^\eta\xi^2,
\end{equation}
where $\eta$ defines the parameter under consideration. 

The parameter $\alpha_1$ deserves special attention because it must embody the intriguing sign dependence
of $g^2F$ on $\nu$ for the SDMs described above. We find the following relation:
\begin{equation}
\label{eq.10}
\alpha_1 = \alpha_1^0 + f_{\alpha_1}(\xi)d.
\end{equation}
The derivation of this result involves a fascinating connection between
dynamical and static radial deformations and will be presented elsewhere 
\cite{elsewhere}. Similar ``family behavior'' \cite{reich} 
and sign oscillations with $\nu$ occur for static deformations 
induced by hydrostatic pressure \cite{elsewhere} and uniaxial stress \cite{gartstein}. 

We now turn to the chirality and diameter dependence of $\alpha_2$ and $\Theta_2$.
These parameters effectively represent a large number of phonon modes that
start to become ``active'' at somewhat higher temperatures (between 350 K and 500 K).
In this energy range, the phonon DOS starts to become substantial, 
and we should expect that characteristics of particular nanotubes (chirality
dependence) should gradually disappear and give rise to universal, graphite-like features. 
Of course, this is only exact in the limit of large $d$, therefore, for the relatively
narrow tubes considered here, it is wise to include some chirality dependence (in the form of 
Eq. \ref{eq.9}) that decays with increasing diameter. We therefore propose the following 
expressions:
\begin{eqnarray}
\label{eq.11}
\Theta_2 = \Theta_2^{\infty} + {\frac {f_{\Theta_2}(\xi)}{d}} \\
\label{eq.12}
\alpha_2 = {\frac {1}{d}}\left(B+{\frac {f_{\alpha_2}(\xi)}{d}}\right).  %
\end{eqnarray}
The overall $1/d$ factor in $\alpha_2$ is simple to understand, 
since $\alpha_1+\alpha_2 = \lim_{T\to\infty}{\frac{dE_g}{dT}}$, i.e.,
$\alpha_2$ is the graphite-like contribution to the limiting slope of $E_g(T)$. 
We expect the high-temperature renormalization of the graphite bands around the Fermi point
to be simply a rescaling of the Fermi velocity. Since the semiconducting SWNTs gaps are
obtained, in a simplified view, by slicing the band structure of graphene, we expect
the corresponding change in the gap to be proportional to the gap itself, i.e., to $1/d$.

Equations (\ref{eq.6}-\ref{eq.12}) provide a ready-to-use
recipe for estimating gap shifts with temperature for any nanotube in this diameter
range \cite{fn3}. The 10 parameters are obtained by best fits: $A=9.45\times10^3$ K, 
$\alpha_1^0 = -1.70 \times 10^{-5}$ eV/K, $\gamma_1^{\alpha_1} = 1.68\times 10^{-6}$ eV/K,
$\gamma_2^{\alpha_1} = 6.47\times 10^{-7}$ eV/K, $\Theta_2^{\infty} = 470$ K, 
$\gamma_1^{\Theta_2} = 1.06\times 10^{3}$ K, $\gamma_2^{\Theta_2} = -5.94\times 10^{-2}$ K,
$B = -4.54 \times 10^{-4}$ eV/K, $\gamma_1^{\alpha_2} = -2.68\times 10^{-3}$ eV/K and
$\gamma_2^{\alpha_2} = -2.23\times 10^{-5}$ eV/K. 

Finally, we address the effects of thermal expansion in $E_g(T)$. Available 
descriptions of thermal expansion of isolated SWNTs seem to be controversial
\cite{kwon,schelling}. We propose that, for low-dimensional structures, one 
should describe thermal expansion effects in the band gap  
in terms of anharmonic changes of internal coordinates (bond lengths and angles), 
rather than lattice constants. The C-C bond
changes quite similarly in all carbon structures (including diamond), although the lattice
constants may behave quite differently \cite{raravikar}. We therefore use the well-established
experimental data for the thermal expansion of diamond \cite{aip}, where the thermal expansion 
of the lattice mirrors closely the thermal expansion of the bond, to estimate the gap shifts
due to anharmonic effects. This leads to very small corrections in our calculated
$E_g(T)$ values. For instance, the additional gap shift at 300 K for a (10,0) tube due to
anharmonic effects is only $-0.2$ meV. We therefore can safely neglect these effects in comparison
with the harmonic contributions.

In conclusion, the calculated temperature dependence of the band gap of semiconducting
carbon nanotubes shows a complex but systematic diameter and chirality dependence. Most gap shifts
at 300 K are negative and small (less than 12 meV) with respect to 0 K. Therefore, temperature
effects should not interfere with $(n,m)$ assignments. Tubes with 
$\nu=1$ have generally larger shifts than $\nu=2$ tubes, and some $\nu=2$ tubes with large
chiral angles even display a non-monotonic $E_g(T)$ curve. All these features are explained
and reproduced by a two-phonon model relation, with diameter- and chirality-dependent parameters.

We acknowledge useful discussions with S. Saito, A. Jorio, R. B. Weisman, J. Wu, P. B. Allen,
J. Lefebvre, P. Schelling and Y.-K. Kwon. RBC acknowledges financial support from the
John Simon Guggenheim Memorial Foundation and Brazilian funding agencies CNPq, FAPERJ,
Instituto de Nanoci{\^e}ncias, FUJB-UFRJ and PRONEX-MCT. Work partially supported by NSF
Grant No. DMR00-87088 and DOE Contract No. DE-AC03-76SF00098.

\bibliography{temp-condmat}

\begin{thebibliography}{35}
\bibitem[{fn3()}]{fn3}
\bibinfo{note}{We have also calculated the temperature shifts for the second
  allowed transition ($E_{22}$). The results can be understood within the same
  framework as $E_g$ (or $E_{11}$), provided that the diameter dependences of
  $\alpha_1$ and $\alpha_2$ are rescaled by a factor of two and the sign factor
  $(-1)^\nu$ is reversed.}
\end{thebibliography}

\end{document}